\documentclass[runningheads,citeauthoryear]{apinv}
\usepackage{epsfig,graphics}
\usepackage{cite}
\usepackage{marvosym}
\usepackage[T2A]{fontenc}
\usepackage[utf8]{inputenc}
\usepackage{float}
\usepackage{tabu}
\usepackage{url}

\begin{document}

\title{Transformation of Pan-STARRS1 \textit{gri} \\ to Stetson BVRI magnitudes. \\Photometry of small bodies observations.}
											
\titlerunning{Connection between BVRI and \textit{gri} ...}
\author{A. Kostov \& T. Bonev}
\authorrunning{A. Kostov \& T. Bonev}
\tocauthor{A. Kostov \& T. Bonev}
\institute{Institute of Astronomy and NAO, Bulgarian Academy of Sciences, BG-1784, Sofia   \newline
	\email{akostov@astro.bas.bg}    }
\papertype{Submitted on 15.05.2017; Accepted on 18.06.2017}	
\maketitle

\begin{abstract}The UBVRI broad band photometric system is widely used in CCD astronomy. There are a lot of sets of standard stars for this photometric system, the Landolt's and Stetson's catalogues being the most precise and reliable. Another photometric system, recently considerably spread in CCD observations is \textit{ugriz}, which originates from the Sloan Digital Sky Survey (SDSS) and has now many variations based on its 5 broad-band filters. One of the photometric systems based on it is The Panoramic Survey Telescope and Rapid Response System (Pan-STARRS). In this paper we compare the BVRI magnitudes in the Stetson catalogue of standard stars with the magnitudes of the corresponding stars in the Pan-STARRS1 (PS1) \textit{grizyw} catalogue. Transformations between these two systems are presented and discussed. An algorithm for data reduction and calibration is developed and its functionality is demonstrated in the magnitude determination of an asteroid.
\end{abstract}

\keywords{Surveys -- Catalogs -- Techniques: photometric -- Comets: general -- Minor planets, asteroids: general}

\section*{Introduction}

One peculiarity of the observations of moving objects is the constantly changing observed field. Every comet or asteroid has unique proper motion, in general absolutely different from the celestial one. In case of fast moving objects, there can be several different observed fields in the same night. Due to that feature, calibration of instrumental magnitudes with photometric standard stars is more difficult in comparison to observations of fixed star fields, and establishing a good set of secondary photometric standards is virtually impossible. It gets evident that in the case of small bodies photometry one needs an all-sky spread standard stars in the BVRI photometric system. Hence the main goal of this work is to obtain reliable relation between an all-sky catalogue PS1 and Stetson BVRI standard stars catalogue (\textbf{Stetson Homogeneous photometry database}\footnote{\url{http://www.cadc-ccda.hia-iha.nrc-cnrc.gc.ca/en/community/STETSON/standards/}\label{SDB}}).

Before the release of PS1 we used to calibrate our photometry with the only suitable broad-band all-sky catalogue - the USNO B1.0 (Monet et al. 2003). Covering the B and R passbands this catalogue shows excellent sky coverage and astrometric precision. Unfortunately USNO B1.0 is characterized by a poor photometric accuracy. Nevertheless we used USNO B1.0 following Kidger (2003). We did some attempts to improve our data calibration by using additional re-calibrations of USNO B1.0 photometry, as suggested by Madsen and Gaensler (2013). The connections between SDSS photometric systems and BVRI (Jordi et al. 2006), were noted but could not be used, because insufficient coverage of the sky. With the release of PS1, this problem is solved, and this motivated us to obtain a relation between PS1 and Stetson’s BVRI photometric standards.
 
\section{Choice of the photometric system}

One of the cornerstones in astronomy is the establishment of reliable celestial objects magnitudes. Selection of a proper photometric system with a suitable set of primary standards is the first, and most important step of obtaining accurate photometric results. There is a large number of photometric systems covering different spectral ranges. One of the most widely used and well established broad-band photometric system is the Johnson-Cousins UBV Kron-Cousins RI (UBVRI). Being one of the first photometric systems, UBVRI underwent a long evolution (Bessell 2005). Nowadays the most used set of filters representing this system is Johnson/Bessell (Bessell 1990). Further in this work we will omit the U band filter for two main reasons: first, there is a lack of data in Stetson's catalogue for U, and second, with very rare exceptions, our observations of small bodies practically did not include the U band.

As was stated above the standardization of stellar photometry presented in this investigation is based on the use of the Stetson’s catalogue of CCD standard stars (Stetson 2000). Initially based on the Landolt (1992) UBVRI photometry, today Stetson’s catalogue comprises a large set of fields covering wide range of declinations. Observations are multy-epoch, extended down to magnitudes fainter than 20 mag. The stellar photometry is homogeneous, precise and regularly updated.

In the last two decades several digital sky surveys use one new photometric system, which may become a new standard in photometric observations. This is the \textit{ugriz} photometric system of the SDSS and it's variations (Fukugita et al. 1996). Pan-STARRS1 survey (Chambers et al 2016) is build on observations obtained by \textit{grizyw} photometric system, and since 12 December 2016 the \textbf{Pan-STARRS1 Catalogue Search}\footnote{\url{http://archive.stsci.edu/panstarrs/search.php}\label{PS1DB}} is public. The survey is multy-epoch, it covers the entire North sky, extends to declinations $-30^{\circ}$, contains wide range of magnitudes with milimagnitude accuracy (Tonry et al. 2012; Magnier et al. 2016a), and give us the opportunity to use it as a secondary standard catalogue. Further in this work we will use only the first half of the entire PS1 photometric system - \textit{gri}. The \textit{zyw} does not correspond to the aim of our task, and as such will be omitted in the transformation equations.

\begin{figure}[h]
\begin{center}
	\centering{\epsfig{file=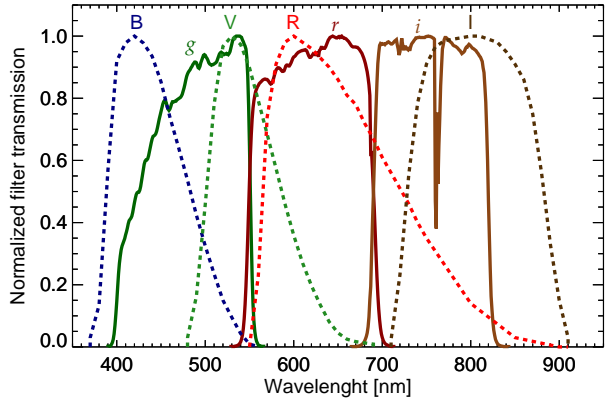, width=0.6\textwidth}}
	\caption[]{Comparison of Johnson/Bessell BVRI and Pan-STARRS1 \textit{gri} filters transmission curves.}
	\label{filters}
\end{center}
\end{figure}

The transmissions of the BVRI (Bessell 1990) and the \textit{gri} filters\footnote{\url{http://iopscience.iop.org/0004-637X/750/2/99/suppdata/apj425122t3_mrt.txt}\label{PS1FS}} (Tonry et al. 2012) are presented in Fig.\ref{filters}\footnote{Further in this work we will use the same colours of the filters as in Fig.\ref{filters} for representing the features connected to BVRI or \textit{gri}\label{CSP}}. The profiles of the filters and the covered spectral ranges are quite different, and there is no simple connection $V\!=\!f(g)$, $R\!=\!f(r)$ or $I\!=\!f(i)$. Due to that we decided to compare both photometric systems in more details using colours instead of magnitudes in the transformation equations.

\newpage
\section{Catalogues Data}

\begin{figure}[h]
\begin{center}
	\hspace{+18 mm}
	\centering{\epsfig{file=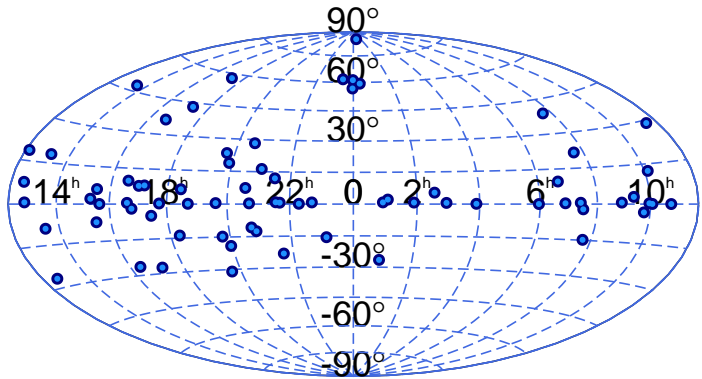, width=0.7\textwidth}}
	\caption[]{Distribution of the used fields across the celestial sphere. The coordinate system of the chart is equatorial.}
	\label{skymap}
\end{center}
\end{figure}

For the purposes of our work 73 fields from Stetson's database\textsuperscript{\ref{SDB}} were selected. In order to obtain relation between both catalogues, only fields for which BVRI photometry is available were used. The sample covers declinations between $-30^{\circ}$ and $+85^{\circ}$ and includes as many different sky patches as possible (Fig.\ref{skymap}).

That initial sample contains 33026 Stetson's standard stars which were cross-identified  with PS1 observations. The  PS1 catalogue search tool\textsuperscript{\ref{PS1DB}}(Flewelling et al. 2016) provide a lot of useful options, such as possibility to upload data file with coordinates and simultaneously identify group of objects (for example all stars from a given observed field can be identified in one run). The image quality of PS1 survey is evaluated by Magnier et al. (2016c) with a median \textit{FWHM}$=\!\![1.31(\textit{g}), 1.19(\textit{r}), 1.11(\textit{i})]''$. The PS1 nominal plate scale of $0.258''$ per pixel (Waters et al. 2016) corresponds to a median PSF of a star about 5 pixels. The astrometric accuracy of the PS1 is $0.022''$ for the right ascension ($\alpha$) and $0.023''$ for the declination ($\delta$) (Magnier et al. 2016c). Although Stetson has not provided individual astrometric errors for the positions of the stars, in his database\textsuperscript{\ref{SDB}} the coordinates of the stars are given with accuracy of $0.01^s$ i.e. $0.15''$ for $\alpha$ and $0.1''$ for $\delta$. In order to produce a homogeneous star list, and to avoid spurious cross-identifications, we used to split the selection process in 3 steps (depending on the search radius used with the PS1 catalogue search tool\textsuperscript{\ref{PS1DB}}) as follows:

\begin{itemize}
\renewcommand{\labelitemi}{$\bullet$}
			\item{\textbf{step1}: search all 33026 Stetson's stars in PS1 catalogue within a search radius $R\!=\!3\!\times\!\sqrt{(0.15^2 + 0.1^2)}\!=\!0.541''$. This yields 24903 cross-identifications, $75.4\%$ of the initial sample.}
\end{itemize}
			
After this first step a manual check of random selected not-cross-identified objects was performed. Stars, with different magnitudes were checked in the \textbf{Image Cutout Server}\footnote{\url{http://ps1images.stsci.edu/cgi-bin/ps1cutouts}\label{PS1IS}} tool of the PS1 archive. This test shows that a significant number of good defined stars, without neighbours or near artefacts, do not match with the first criterion. The possible reasons for that pitfall are explained in \textbf{PanSTARRS Search Help}\footnote{\url{http://archive.stsci.edu/panstarrs/help/search_help.html}\label{PS1SH}} under the description of the \textbf{Radius}: \textit{'You should be careful about giving too restrictive a search radius since (for some missions) the coordinates of the object were given by the Guest Observer, and may not reflect the precise pointing of the instrument at the time of the observation'}. To extend our sample we applied the next 2 steps:

\begin{itemize}
\renewcommand{\labelitemi}{$\bullet$}
			\item{\textbf{step2}: search all not-cross-identified 8123 stars within $R\!=\!0.655''$. In order to account for possible shifts of stars positions, related to the quality of the image, we extended the search aperture to the median \textit{FWHM} of PS1. This second search adds more 6102 stars to the list of cross-identified stars.}
			\item{\textbf{step3}: search all remaining 2021 stars with $R\!=\!2.62''$, i.e $2\times$\textit{FWHM}$(g)$. This generous search radius gives 2005 new matches. To prevent the sample from nonstar identifications (see Fig. 3 in Magnier et al. 2016b) we decided to use only these with angular separation $\rho\!\le\!1.0''$, which are 1768 stars.}
\end{itemize}

\begin{figure}[h]
\begin{center}
	\centering{\epsfig{file=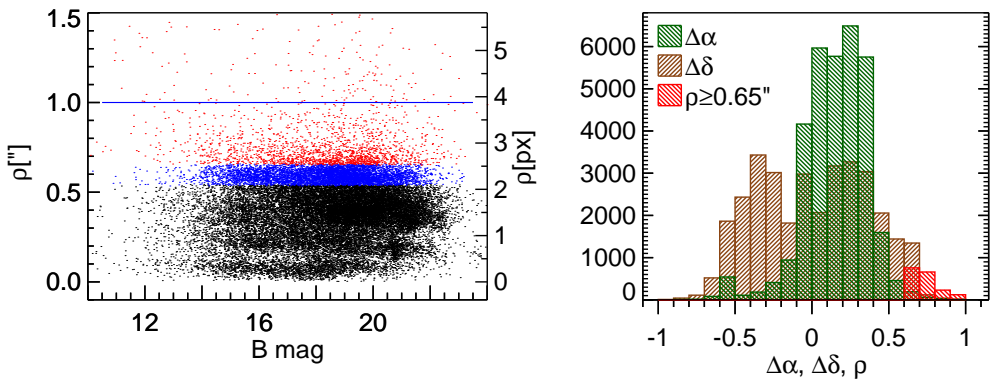, width=0.99\textwidth}}
	\caption[]{Angular separations (left) and differences in $\alpha$, $\delta$ (right) for 32773 cross-identified stars.}
	\label{astrometry}
\end{center}
\end{figure}

The final cross-identification list contains 32773 stars, which is $99.2\%$ of the initial Stetson's data. The angular distance between the Stetson's stars and their PS1 counterparts is a good measure for the quality of the cross-identification. Fig.\ref{astrometry} shows the distribution of these angular distances. In the left panel of Fig.\ref{astrometry} the left ordinate is in arcseconds, and the right one shows the corresponding distances in pixels. The black dots are the stars with $\rho\!\le\!0.541''$, the blue dots are the stars within $\rho\!=\![0.541,0.655]''$, and the red dots are the stars within $\rho\!=\![0.655,2.62]''$, the upper limit of the latter being outside of the ordinate range. Below the blue line ($1.0''$) are all stars which passed the astrometric criteria. For them, in the right panel of this figure, the distributions of the differences in $\alpha$, $\delta$ are shown. In this histogram plot is also shown the distribution of $\rho$ for the set of 1768 stars, selected by using the selection criterion described in \textbf{step3}. As seen, the number of stars drops rapidly, and for $\rho > 1.0''$ the number of omitted  stars is as low as 237.

After finishing the astrometric cross-identifications, we set 3 more conditions, depending on photometric data of the selected stars:
\begin{itemize}
\renewcommand{\labelitemi}{$\bullet$}
		\item{All stars must have Stetson BVRI and PS1 \textit{gri} magnitudes. This criterion removes 612 stars from the list.}
		\item{According to Stetson (2000) his database\textsuperscript{\ref{SDB}} satisfies the following criterion: \textit{'at least five independent observations under photometric conditions and standard errors of the mean magnitude smaller than 0.02 mag'}. We applied the second part of this requirement to the cross-identified PS1 stars, i.e. we removed from the final list selected set all stars having photometry errors greater than 0.02 magnitudes. This criterion removes 8198 stars.}
				\item{All stars are restricted to magnitude $<19$ mag.(see the note at the end of section~\ref{sec:TE} for more detailed explanation of this criterion). Application of this criterion removes 14971 stars from the list.} \end{itemize}

Note that all three photometric criteria were applied simultaneously, and the final amount of removed stars from the cross-identification list is 17699. Because some of the stars met more than one criterion, the reduction of the list is not a simple sum of the removals of each criterion. The final list of stars which we used to derive the colour transformations between PS1 \textit{gri} and Stetson’s BVRI catalogue was reduced to N = 15074. These standard stars are distributed in 68 Stetson's fields. Table \ref{table1} contains the names of the fields (column 1) as given in the database\textsuperscript{\ref{SDB}}, the central coordinates of the fields for epoch (2000.0) (columns 2-7) and the number of stars for each particular field (column 8) used to derive the colour transformations.

\begin{table}[H]
\scriptsize
  \begin{center}
  \caption{Stetson standard fields used for comparison with PS1 catalogue}
  \setlength{\extrarowheight}{1.5pt}
	\begin{tabu}{lccr|lccr|lccr}
	\hline
Field  		& RA 					& DEC 			& N		& Field 	& RA 					& DEC 			& N & Field  		& RA 					& DEC 			& N		 \\
  \hline

   NGC6809 & 19 40.0 & -30 57.7 &  130 &     L98 & 06 52.0 & -00 20.6 & 1502 & UGC10743 & 17 11.3 & +07 56.0 &  203   \\
 Barnard59 & 17 11.7 & -27 24.9 &   61 &    L101 & 09 57.3 & -00 17.3 &  337 &  NGC2264 & 06 41.1 & +09 38.3 &   74   \\
   NGC4590 & 12 39.5 & -26 44.6 &   31 &    L109 & 17 44.7 & -00 15.4 &  198 &   PG1633 & 16 35.5 & +09 47.0 &  268   \\
    NGC288 & 00 52.5 & -26 27.7 &  173 &  PG2213 & 22 17.4 & -00 08.8 &  111 &  NGC7078 & 21 29.6 & +11 59.3 &  113   \\
   NGC6121 & 16 23.3 & -26 25.3 &  166 &    Pal3 & 10 04.2 & -00 03.0 &   43 &     LeoI & 10 08.3 & +12 23.1 &  154   \\
   NGC7099 & 21 40.3 & -23 10.5 &  166 &  PG1047 & 10 50.0 & -00 01.0 &   18 &  NGC7006 & 21 01.4 & +16 17.1 &  188   \\
      133P & 19 57.5 & -19 07.7 &   37 &    L107 & 15 38.5 & +00 02.9 &  650 &  NGC5024 & 13 12.9 & +18 10.2 &  207   \\
   NGC7492 & 23 08.1 & -15 36.7 &   29 &     L95 & 03 53.6 & +00 07.3 &  225 &  NGC4147 & 12 10.2 & +18 31.4 &   77   \\
   NGC2437 & 07 41.8 & -14 48.9 &  246 &    L110 & 18 42.8 & +00 08.3 &  322 &  NGC6838 & 19 53.9 & +18 46.7 &   35   \\
   NGC6822 & 19 44.7 & -14 46.8 &  311 &    L112 & 20 42.6 & +00 19.7 &   37 &  NGC2420 & 07 38.4 & +21 34.4 &  145   \\
 Messier16 & 18 18.6 & -13 50.9 &  246 &    L111 & 19 37.7 & +00 27.0 &   77 &  NGC6823 & 19 43.2 & +23 18.0 &   37   \\
   NGC6981 & 20 53.5 & -12 32.2 &   48 &    L113 & 21 41.1 & +00 39.2 &  672 &  NGC6940 & 20 34.6 & +28 16.1 &  127   \\
     MarkA & 20 44.7 & -10 45.9 &  822 &    L114 & 22 41.9 & +00 46.8 &    2 &     Pal4 & 11 29.3 & +28 58.4 &    8   \\
    PG1323 & 13 26.2 & -08 58.4 &  528 &     L92 & 00 55.5 & +00 47.7 &  221 &  NGC6205 & 16 41.7 & +36 27.6 &  187   \\
    PG1525 & 15 28.2 & -07 15.7 &  384 &  IC1613 & 01 04.8 & +02 07.1 &   86 &  NGC2419 & 07 37.9 & +38 54.5 &  127   \\
   NGC6366 & 17 27.4 & -05 01.2 &   64 & NGC5904 & 15 18.2 & +02 08.1 & 1105 &  NGC6341 & 17 17.1 & +43 08.2 &  585   \\
    PG0942 & 09 45.2 & -03 08.5 &   22 &  PG0918 & 09 21.7 & +02 46.5 &  104 &  NGC7789 & 23 57.4 & +56 42.5 &   19   \\
     Ru152 & 07 29.8 & -02 10.3 &  137 &  PG0231 & 02 33.7 & +05 18.7 &   26 &    Draco & 17 19.4 & +57 55.5 &   51   \\
   NGC6218 & 16 47.2 & -01 56.9 &  172 &  PG1528 & 15 30.8 & +06 01.3 &   19 &     IC10 & 00 20.4 & +59 17.6 &   48   \\
   NGC7089 & 21 33.4 & -00 50.7 &  437 & NGC6633 & 18 27.3 & +06 31.8 &   46 &  NGC7790 & 23 58.4 & +61 13.0 &  235   \\
      L104 & 12 42.1 & -00 35.6 &  298 & NGC6934 & 20 34.2 & +07 24.2 &   94 &  NGC7654 & 23 24.9 & +61 38.9 &  116   \\
     Ru149 & 07 24.3 & -00 32.9 & 1429 & NGC4526 & 12 34.1 & +07 42.0 &    6 &   NGC188 & 00 47.5 & +85 16.2 &  172   \\
      L108 & 16 37.9 & -00 30.1 &    9 &  PG1657 & 16 59.6 & +07 42.4 &   51 &          &         &          &        \\
	\hline
  \end{tabu}
  \label{table1}
  \end{center}
\end{table}

\section{Transformation equations}
\label{sec:TE}
\begin{figure}[H]
\begin{center}
	\centering{\epsfig{file=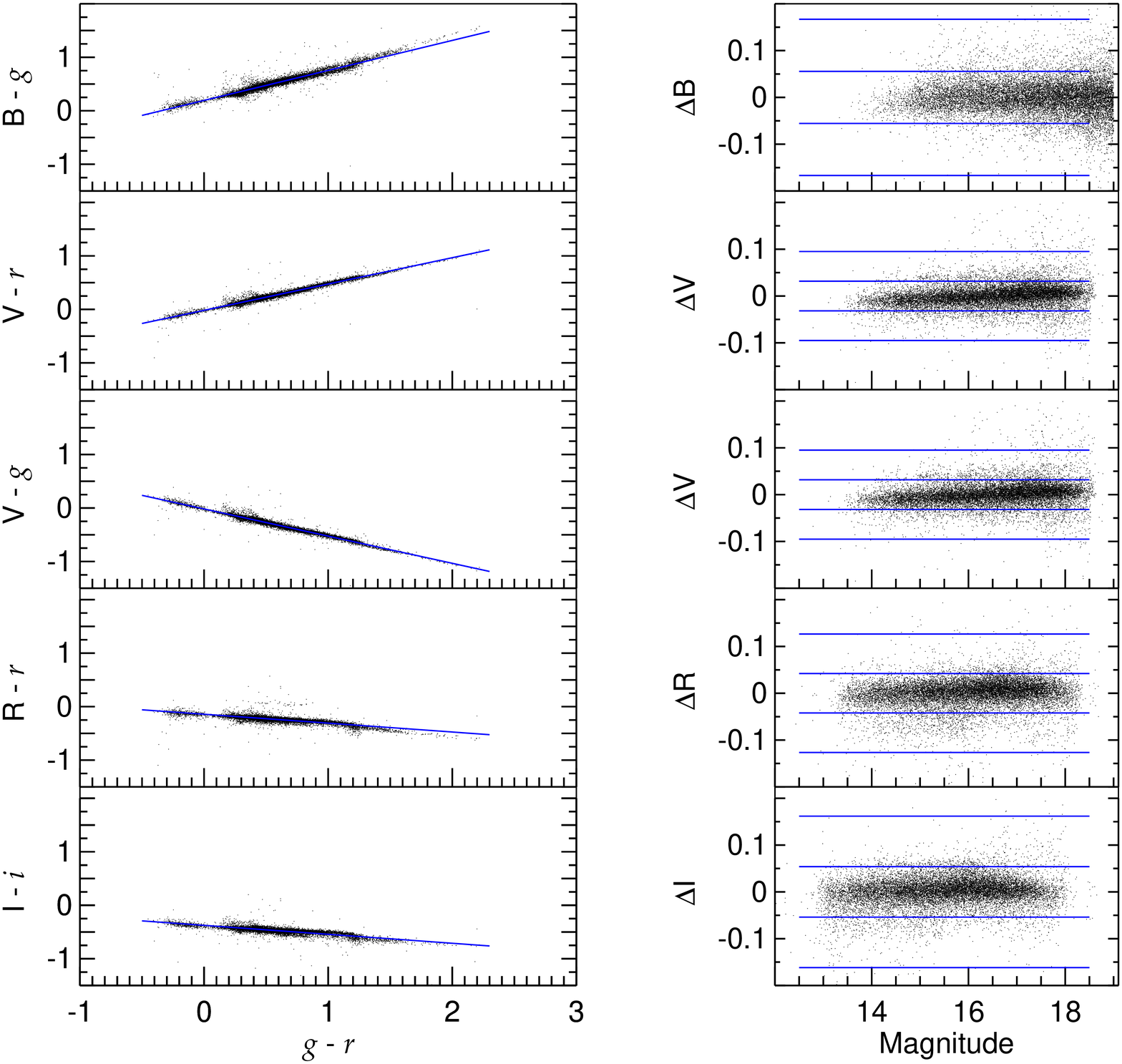, width=0.99\textwidth}}
	\hspace{10 mm}
  \caption[]{Left: Transformations between Stetson’s and PS1 standard stars magnitudes as a function of $(g-r)$ colour indices. Right: Corresponding differences between transformed and Stetson's magnitudes versus magnitude.}
  \label{gr}
\end{center}
\end{figure}

\begin{figure}[H]
\begin{center}
	\centering{\epsfig{file=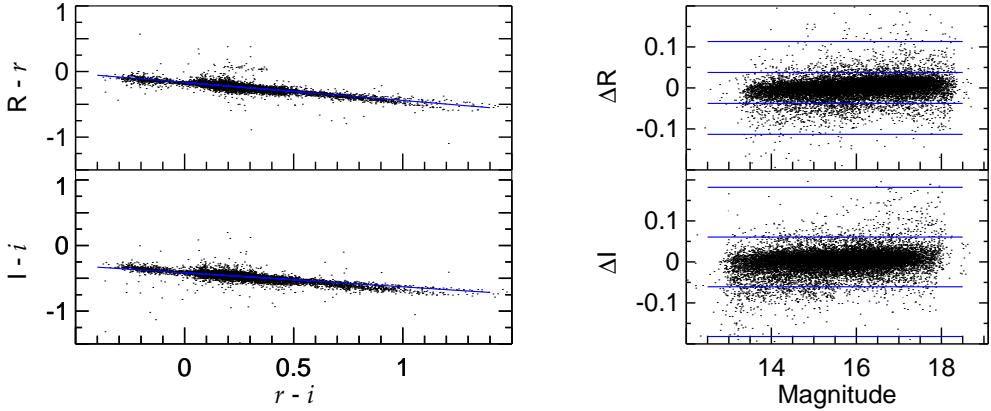, width=0.99\textwidth}}
	\hspace{10 mm}
  \caption[]{Left: Transformations between Stetson’s and PS1 standard stars magnitudes as a function of $(r-i)$ colour indices. Right: Corresponding differences between transformed and Stetson's magnitudes versus magnitude.}
  \label{ri}
\end{center}
\end{figure}

The comparison between PS1 \textit{gri} and Stetson’s BVRI magnitudes could be described with the following seven relations:
\begin{equation}
B-g=C_0+C_1\times(g-r)  \quad\quad\quad   B-g=D_0+D_1\times(g-r)+D_2\times(g-r)^2 \quad\quad
\label{eq:bggr}
\end{equation}
\vspace{-5 mm}
\begin{equation}
V-g=C_0+C_1\times(g-r)  \quad\quad\quad   V-g=D_0+D_1\times(g-r)+D_2\times(g-r)^2 \quad\quad
\label{eq:vggr}
\end{equation}
\begin{equation}
V-r=C_0+C_1\times(g-r)  \quad\quad\quad   V-r=D_0+D_1\times(g-r)+D_2\times(g-r)^2 \quad\quad
\label{eq:vrgr}
\end{equation}
\begin{equation}
R-r=C_0+C_1\times(g-r)  \quad\quad\quad   R-r=D_0+D_1\times(g-r)+D_2\times(g-r)^2 \quad\quad
\label{eq:rrgr}
\end{equation}
\begin{equation}
I-i=C_0+C_1\times(g-r)  \quad\quad\quad   I-i=D_0+D_1\times(g-r)+D_2\times(g-r)^2 \quad\quad
\label{eq:iigr}
\end{equation}
\begin{equation}
R-r=C_0+C_1\times(r-i)  \quad\quad\quad   R-r=D_0+D_1\times(r-i)+D_2\times(r-i)^2 \quad\quad
\label{eq:rrri}
\end{equation}
\begin{equation}
I-i=C_0+C_1\times(r-i)  \quad\quad\quad   I-i=D_0+D_1\times(r-i)+D_2\times(r-i)^2 \quad\quad
\label{eq:iiri}
\end{equation}

where ($C_i$) and ($D_i$) are the coefficients in the first and second order approximations, respectively. The resulting coefficients derived from fits to the data are summarised in Table \ref{equation_results}. Here the first and the second columns contains the coefficients of the linear fit, followed by the standard deviation of the fit, $\sigma$. The columns 4-6 contain the coefficients of the quadratic fit, and in column 7 is the standard deviation of this fit, $\sigma_{2}$. The uncertainties are placed under the coefficients. In most of the cases the standard deviation of the linear fit is sufficiently small, and the coefficients ($C_0, C_1$), derived from the first order link between both photometric systems can be used with confidence. We observe a slight deviation from the liner fits around the boundaries of the used colours. The linear fit is correct for colours $(g-r)$ in the range [-0.5, 1.5], and for $(r-i)$ in the range [-0.4, 1.0] (see Fig. \ref{gr} and Fig. \ref{ri}), i.e. practically for more than $80\%$ of the colour range, excluding stars with extremely red colour indices.

\begin{table}[H]
  \begin{center}
  \caption{Transformation coefficients derived from the linear $(C_i)$ and second order $(D_i)$ fits of the data described by equations (\ref{eq:bggr})$\div$(\ref{eq:iiri})}
  \setlength{\extrarowheight}{3.0pt}
	\setlength{\tabcolsep}{5pt}
	\begin{tabu}{ccc|cccc}
	\hline
$C_0$ & $C_1$ & $\sigma$ & $D_0$ & $D_1$ & $D_2$ & $\sigma_{2}$ \\

	\hline

\multicolumn{7}{c}{$B-g = f(g-r)$} \\
      0.194  &      0.561 & 0.056 &      0.199 &      0.540 &      0.016 & 0.056 \\
 $(\pm0.001)$  & $(\pm0.002)$ &       & $(\pm0.001)$ & $(\pm0.004)$ & $(\pm0.003)$ &        \\

\hline

\multicolumn{7}{c}{$V-g = f(g-r)$} \\
      -0.017  &      -0.508 & 0.032 &      -0.020 &      -0.498 &      -0.008 & 0.032 \\
 $(\pm0.001)$  & $(\pm0.001)$ &       & $(\pm0.001)$ & $(\pm0.002)$ & $(\pm0.002)$ &        \\

\hline

\multicolumn{7}{c}{$V-r = f(g-r)$} \\
      -0.017  &      0.492 & 0.032 &      -0.020 &      0.502 &      -0.008 & 0.032 \\
 $(\pm0.001)$  & $(\pm0.001)$ &       & $(\pm0.001)$ & $(\pm0.002)$ & $(\pm0.002)$ &        \\

\hline

\multicolumn{7}{c}{$R-r = f(g-r)$} \\
      -0.142  &      -0.166 & 0.042 &      -0.163 &      -0.086 &      -0.061 & 0.041 \\
 $(\pm0.001)$  & $(\pm0.001)$ &       & $(\pm0.001)$ & $(\pm0.003)$ & $(\pm0.002)$ &        \\

\hline

\multicolumn{7}{c}{$R-r = f(r-i)$} \\
      -0.166  &      -0.275 & 0.038 &      -0.172 &      -0.221 &      -0.081 & 0.036 \\
 $(\pm0.000)$  & $(\pm0.002)$ &       & $(\pm0.000)$ & $(\pm0.002)$ & $(\pm0.002)$ &        \\

\hline

\multicolumn{7}{c}{$I-i = f(g-r)$} \\
      -0.376  &      -0.167 & 0.054 &      -0.387 &      -0.123 &      -0.034 & 0.054 \\
 $(\pm0.001)$  & $(\pm0.001)$ &       & $(\pm0.001)$ & $(\pm0.004)$ & $(\pm0.003)$ &        \\

\hline

\multicolumn{7}{c}{$I-i = f(r-i)$} \\
      -0.416  &      -0.214 & 0.061 &      -0.433 &      -0.040 &      -0.263 & 0.048 \\
 $(\pm0.001)$  & $(\pm0.003)$ &       & $(\pm0.001)$ & $(\pm0.003)$ & $(\pm0.003)$ &        \\

\hline
	\end{tabu}
  \label{equation_results}
  \end{center}
\end{table}

In most of the cases the coefficients of the quadratic term in the second order fits have values lower than the standard deviations of the fits (both, first and second order). Here we have two exceptions: the $(r-i)$ colour equations (\ref{eq:rrri})$\div$(\ref{eq:iiri}) have significantly higher quadratic terms, and in this cases, the quadratic fits should be used for red stars, where these distributions are described better in comparison to the linear fit (see Fig. \ref{gr}, Fig. \ref{ri}). In general, quadratic solutions should be used for the regions close to the boundaries of the used colours.

The results are in very good agreement with Tonry et al. (2012) who studied equations (\ref{eq:bggr})$\div$(\ref{eq:iigr}), using synthetic magnitudes with colours ranging from the colour of the Sun to that of Vega, i.e. their transformations don't cover the range $(g-r)\!=\![1.5, 2.5]$, where we notice deviations from the linear fits.

Having the coefficients in the transformation equations we can visualize the results by plotting them against the corresponding colour. The transformations were divided into 2 groups, depending on the colour: $(g-r)$ colour group (equations (\ref{eq:bggr})$\div$(\ref{eq:iigr})), which are presented in Fig. \ref{gr}, and $(r-i)$ colour group (equations (\ref{eq:rrri})$\div$(\ref{eq:iiri})), which are presented in Fig. \ref{ri}. On these similar figures on the left side are shown colour-colour plots for 15074 cross-identified stars. Here the corresponding linear fits are plotted with blue lines. The right columns of these figures contain the deviations of the calculated values from the Stetson's catalogue magnitudes presented as function of magnitude, $\Delta m\!=\!m_{catalogue}\!-\!m_{calculated}$.

The derived differences don't show significant bias from the zero slope. Only in the case of V filter equations, (\ref{eq:vggr}) and (\ref{eq:vrgr}), an increase of the deviations is visible, but $\sim99\%$ of all $\Delta V$ values are inside of the $\pm 3\sigma$ range. Here the blue lines represent $\pm \sigma$ and $\pm 3\sigma$ of the corresponding transformation fit. The majority of the differences $\Delta m$ are within $\pm 3\sigma$ of the fit.

\begin{figure}[H]
\begin{center}
	\centering{\epsfig{file=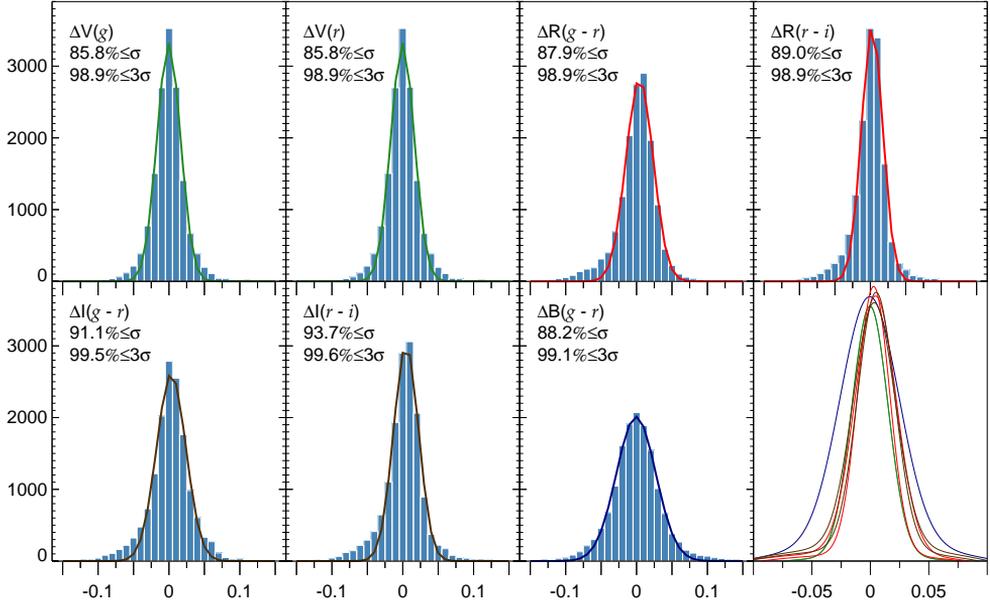, width=0.99\textwidth}}
	\caption[]{Distributions of the differences “calculated values - Stetson's magnitudes”, $\Delta m\!=\!m_{catalogue}\!-\!m_{calculated}$ (corresponds to the data presented in the right panels of Fig. \ref{gr} and Fig. \ref{ri}).}
	\label{histogram}
\end{center}
\end{figure}

Histogram distributions of the differences, $\Delta m$ are shown in Fig. \ref{histogram}. Each histogram plot from left to right represents the distribution of $\Delta m$ derived from equations (\ref{eq:bggr})$\div$(\ref{eq:iiri}). For all transformation equations up to $\sim86\%$ of the stars differ from the catalogue magnitudes within $\pm\sigma$, and up to $\sim98\%$ of the stars deviate from their catalogue magnitudes within $\pm 3\sigma$. Here $\sigma$ is the standard deviation of the fits presented in the left panels of Fig. \ref{gr} and Fig. \ref{ri} (see also Table \ref{equation_results}, column 3, $\sigma$). Each of the histograms was fitted with a Gaussian function. The derived distributions are highly symmetrical around the zero point, close to Gaussian functions, with exception of their higher wings and maxima. There is a slight bias of the wings in negative direction for all distributions, pointing that the calculated magnitudes of stars with greater errors are systematically greater (fainter) compared to the catalogue's. The distributions are characterized by similar maxima - all around 3000, with exception of $\Delta B(g-r)$ (i.e. equation (\ref{eq:bggr})), which has a lower maximum (2033) and wider wings (note the more 'noisier' scatter of the data in the top right plot of Fig. \ref{gr}). The lower right plot in Fig. \ref{histogram} shows the Gaussian fits to each of the 7 histograms, normalized to their maxima. Here only the blue Gaussian ($\Delta B(g-r)$) differs from all other, being broader. We do not observe any humps or tails in the distributions, which is an indication of coherence and homogeneity between the compared photometric systems.

\begin{figure}[H]
\begin{center}
	\centering{\epsfig{file=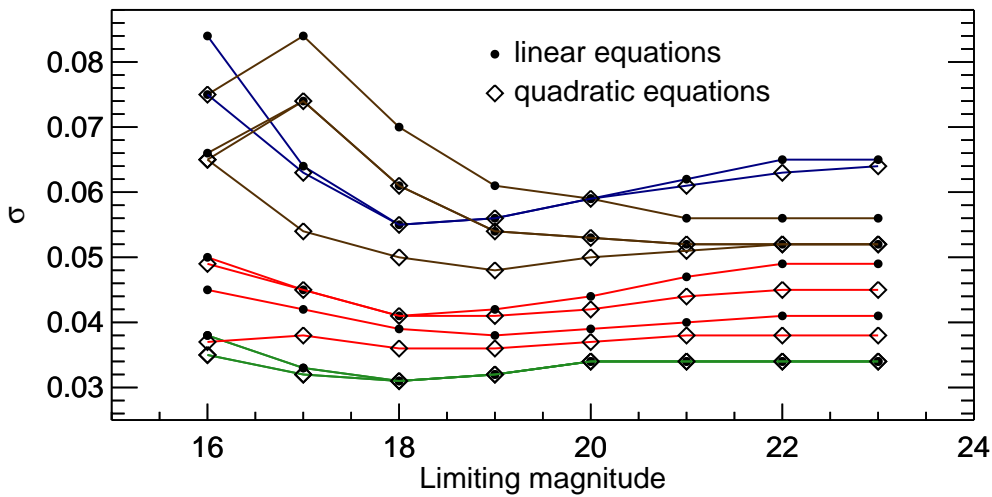, width=0.99\textwidth}}
	\caption[]{Standard deviations, $\sigma$ of the fits, described by the transformation equations (\ref{eq:bggr})$\div$(\ref{eq:iiri}) depending on the limiting magnitude, greater and smaller than 19 mag (photometry criterion 3).}
	\label{errplot}
\end{center}
\end{figure}

Note:	There were 3 photometric limits set to the cross-identified stars. The limiting magnitude is the most questionable one. To check the selection of a  reasonable limiting magnitude we calculated the differences between the catalogue magnitudes and the results of the transformation equations, and analysed their dependence on the number of stars, which results from the magnitude limit. As a measure of the quality of the transformation we used the standard deviation of the linear fit, $\sigma$. Fig. \ref{errplot} presents the standard deviations of linear and quadratic fits for each 7 transformation equations, for 8 different limiting magnitudes, ranging from 16 mag to 23 mag. Restricting the sample to brighter stars, or extending it to the maximum of the magnitude range, raises the errors and noise in the resulting transformations. Minimum of the 1-sigma uncertainties is visible in the magnitude range 18-19 mag. Limiting magnitude of 19 mag gives more populated and thus more representative sample of stars, and therefore it was selected as one of the photometric criteria.

\section{Verification with original observations}

The equations, presented in the previous section, show the relation between two standard stars catalogues, PS1 and Stetson. In order to test the reliability of these catalogue-to-catalogue relations we observed three of the Stetson's fields listed in Table \ref{table1}: L107, NGC5904 and NGC6341. The observations were obtained with the Schmidt telescope of the National astronomical observatory Rozhen, Bulgaria\footnote{\url{http://nao-rozhen.org/telescopes/fr17.html}}.

\begin{figure}[H]
\begin{center}
	\centering{\epsfig{file=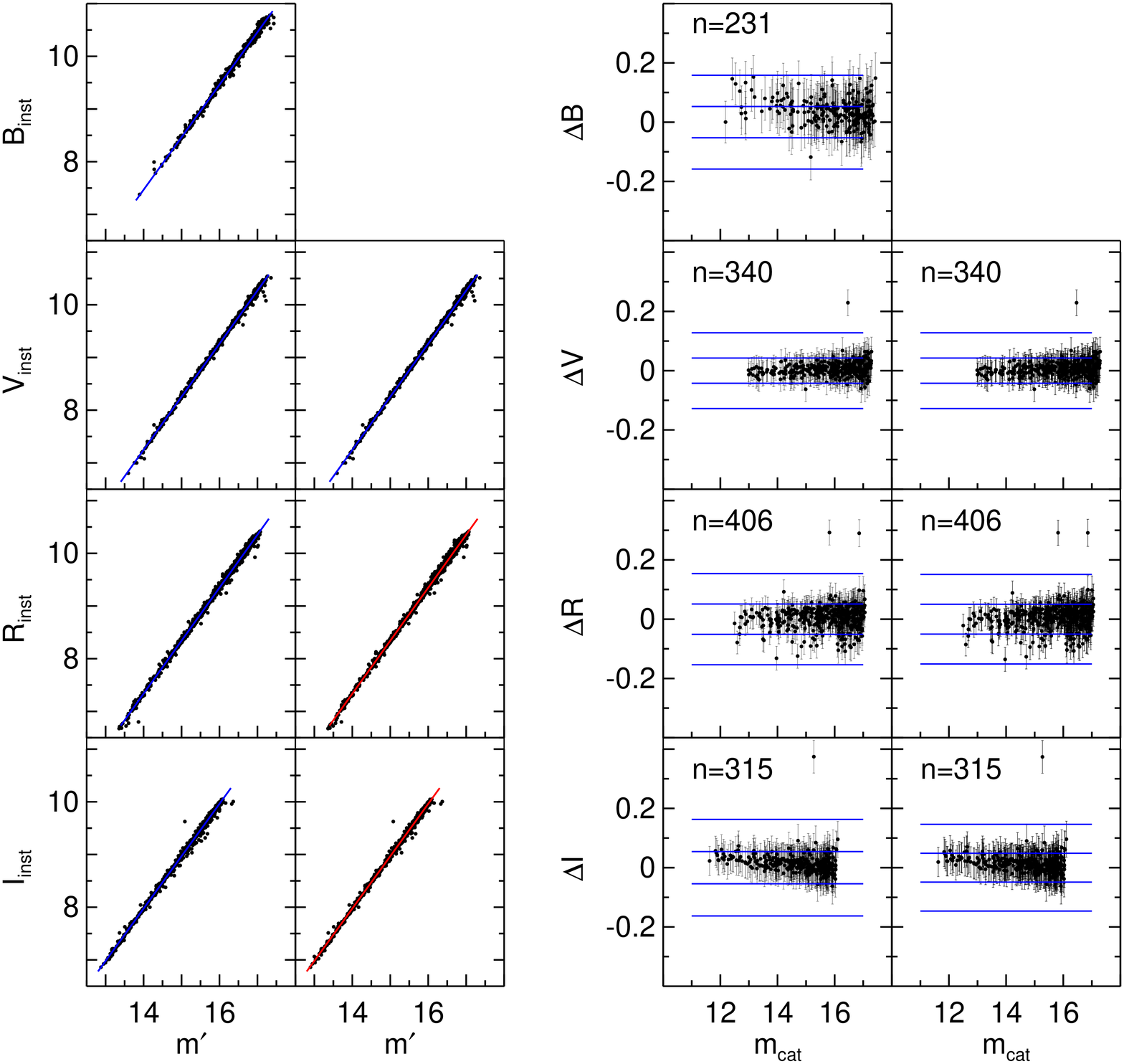, width=0.99\textwidth}}
	\caption[]{Left: Instrumental magnitudes compared to transformed $B^\prime V^\prime R^\prime I^\prime$ magnitudes for all 7 equations. Right: Deviations of the calculated Stetson's magnitudes relative to their catalogue values $\Delta m\!=\!m_{catalogue}\!-\!m_{calculated}$. Stetson's field L107}
	\label{L107}
\end{center}
\end{figure}

\begin{figure}[H]
\begin{center}
	\centering{\epsfig{file=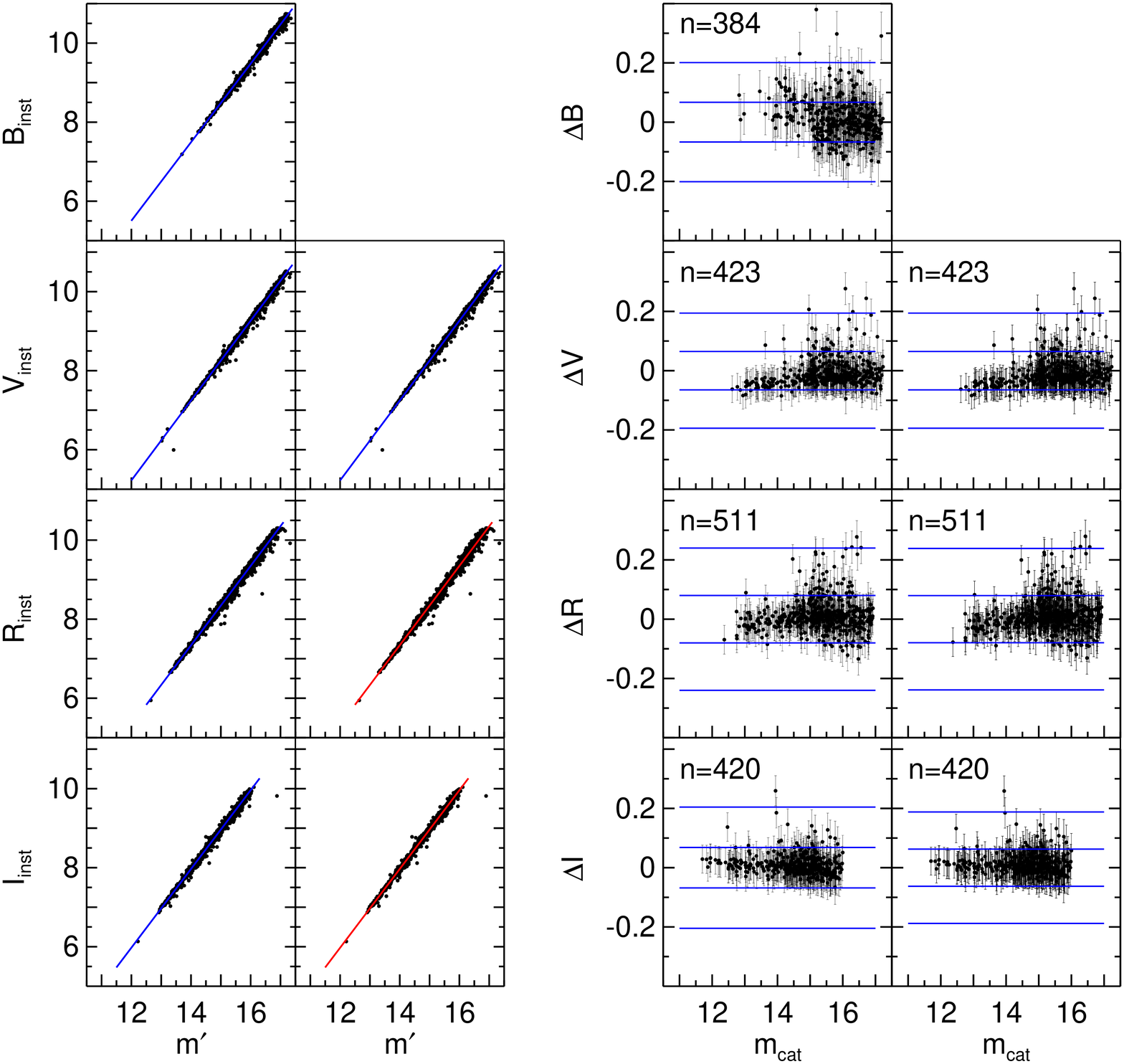, width=0.99\textwidth}}
	\caption[]{Same as Fig. \ref{L107}, but for Stetson's field NGC5904}
	\label{NGC5904}
\end{center}
\end{figure}

\begin{figure}[H]
\begin{center}
	\centering{\epsfig{file=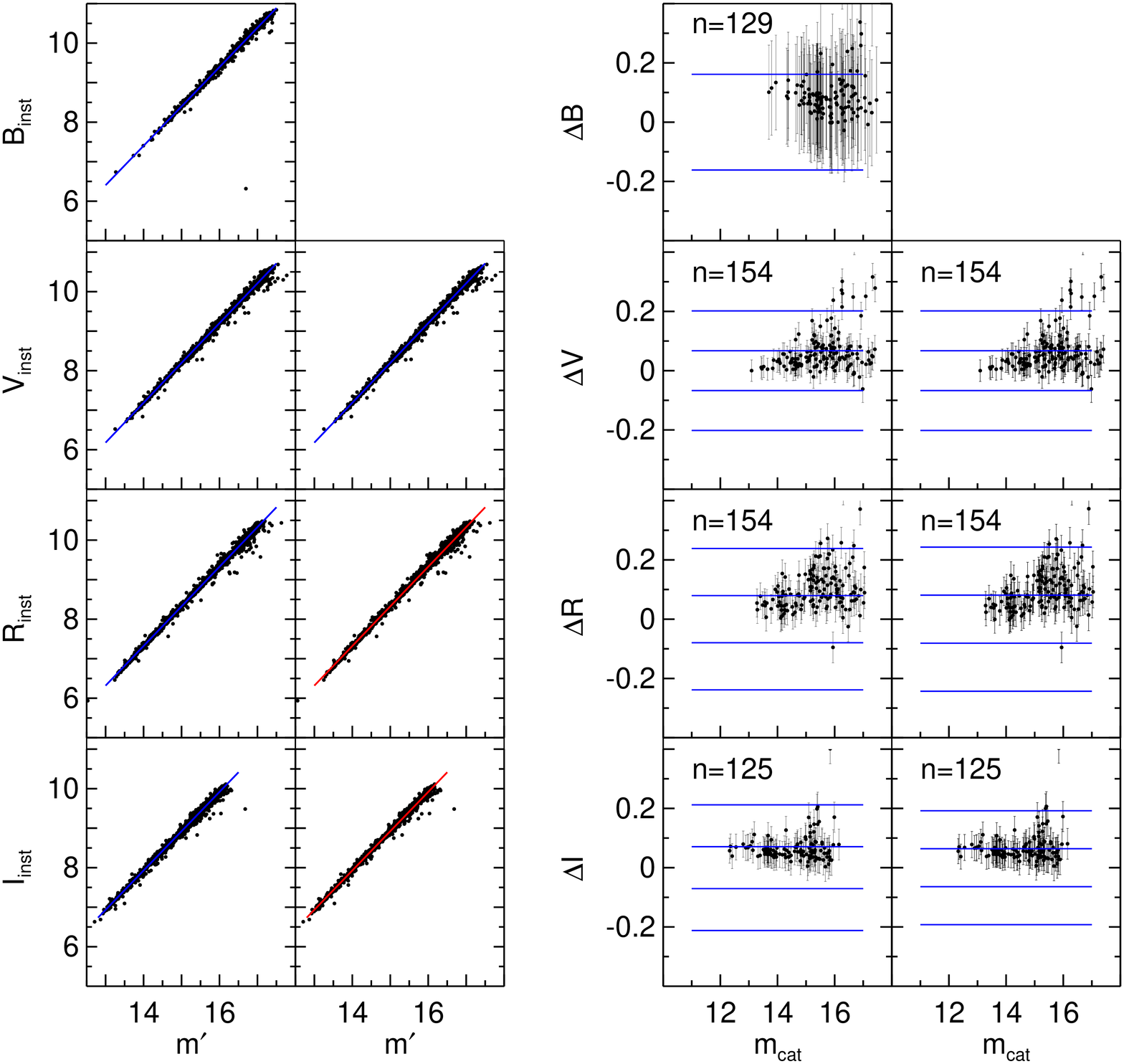, width=0.99\textwidth}}
	\caption[]{Same as Fig. \ref{L107}, but for Stetson's field NGC6341}
	\label{NGC6341}
\end{center}
\end{figure}

The standard procedures of dark subtraction, and flat field divisions were applied to the observational data. After that the initial astrometric solutions for all images were found and all point sources were detected. For each object two sets of coordinates were derived - (X, Y) and ($\alpha, \delta$). All objects were searched (trough their $\alpha, \delta$ coordinates) in the PS1 catalogue search tool\textsuperscript{\ref{PS1DB}}, and a cross-identification list for each image was made. Aperture photometry of all objects was performed. Then the transformations (\ref{eq:bggr})$\div$(\ref{eq:iiri}), with their coefficients (Table \ref{equation_results}), were applied to the PS1 catalogue magnitudes \textit{gri}, to obtain the transformed magnitudes $m^\prime$ - $B^\prime V^\prime R^\prime I^\prime$ (the prime-quantities were introduced as an intermediate designation, preserving BVRI for the final result). The relation between instrumental magnitudes and those calculated by using the transformations (\ref{eq:bggr})$\div$(\ref{eq:iiri}) can be expressed by a simple linear relation -

\begin{equation}
m_{inst} = Z + S\times m^\prime,
\label{eq:zs}
\end{equation}

where Z is the zero point, and S is the slope. Applying a linear fit to the data, the slope and zero point were derived. The results of the fitting procedure are shown in Fig. \ref{L107}, Fig. \ref{NGC5904} and Fig. \ref{NGC6341}. The left column of plots in these figures represents the instrumental magnitudes against the transformed $B^\prime V^\prime R^\prime I^\prime$ magnitudes. These fits are plotted in two colours: blue lines represent $(g-r)$ colour group (equations (\ref{eq:bggr})$\div$(\ref{eq:iigr})), and the members of the $(r-i)$ colour group are red (equations (\ref{eq:rrri}) and (\ref{eq:iiri})). The connections between instrumental and $B^\prime V^\prime R^\prime I^\prime$ magnitudes are well described by linear fits with slopes close to 1.0 and zero points in the interval [-6.92,-5.94]. The  photometric coefficients for all test images are presented in tables  \ref{photL107}, \ref{photNGC5904}, and \ref{photNGC6341}. In these tables the four rows contain N - the number of stars used to estimate the zero point and the slope, S - the slope, Z - the photometric zero point for each image, and $\sigma$ - the standard deviation of the linear fit. The numbers under the values of S and Z are their 1-sigma uncertainty estimates.

With the derived zero point and slope, the standard magnitude of any object can be calculated:

\begin{equation}
m_{calculated} = (m_{inst} - Z)/S
\label{eq:mcalc}
\end{equation}

In this test we consider the Stetson's standards as 'unknown' objects. The instrumental magnitudes of these 'unknown' objects are used now together with the derived photometric coefficients, $Z$ and $S$, to obtain their calculated magnitudes, ($m_{calculated}$). Comparing them to the catalogue magnitudes for each 'unknown' Stetson standard we can verify the reliability of the transformation equations (\ref{eq:bggr})$\div$(\ref{eq:iiri}). In the right plots of figures \ref{L107}, \ref{NGC5904}, and \ref{NGC6341} the deviations between the catalogue and the calculated Stetson's magnitudes $\Delta m\!=\!m_{catalogue}\!-\!m_{calculated}$ are shown. Here the blue lines represent $\pm \sigma$ and $\pm 3\sigma$ of the transformation plots. Symbol \textit{n} in the panels stays for the numbers of the checked Stetson standards. Deviation's scatter grows up in the direction of fainter magnitudes, where the instrumental values are derived with higher errors. Nevertheless the majority of the differences are close to the zero line, and within the $3\sigma$ limits of the fits, indicating that the transformations are coherent. The deviations of the magnitudes for NGC6341 are shifted from the zero line in positive direction, indicating overestimation of the instrumental magnitudes. The bias is around 1-sigma and rises significantly for fainter stars. One possible explanation of this result is the very low metallicity of the NGC6341 (M92) [Fe/H] = -2.31 (Harris 1996)\footnote{\url{http://physwww.physics.mcmaster.ca/~harris/mwgc.dat}\label{HGCC}}. The metal-poor Globular Clusters, like NGC6341, carry Population II stars, and this can affect the transformation equations, as assumed by Jordi et al. (2006). For the other two Stetson's fields, as for the general sample, we do not observe any effects of the metallicity on the transformation equations.

\begin{table}[H]
  \begin{center}
  \caption{Photometric coefficients for L107}
  \setlength{\extrarowheight}{5pt}
	\setlength{\tabcolsep}{5pt}
	\begin{tabu}{c|c|cc|cc|cc}
	\hline
	& \multicolumn{1}{c|}{B} & \multicolumn{2}{c|}{V} & \multicolumn{2}{c|}{R} & \multicolumn{2}{c}{I} \\
	\hline

N	&	454	& \multicolumn{2}{c|}{677}	& \multicolumn{2}{c|}{903}	& \multicolumn{2}{c}{696} \\
S	&	0.997					&	1.008					& 1.008 				&	1.006					&	1.008					&	0.994					&	0.995					\\
	&	($\pm0.003$)	&	($\pm0.002$)	&	($\pm0.002$)	&	($\pm0.002$)	&	($\pm0.002$)	&	($\pm0.003$)	&	($\pm0.002$)	\\
Z	& -6.50					&	-6.86					& -6.86 				&	-6.75					&	-6.78					&	-5.94					&	-5.95					\\
	&	($\pm0.06$)		&	($\pm0.03$)		&	($\pm0.03$)		&	($\pm0.03$)		&	($\pm0.03$)		&	($\pm0.04$)		&	($\pm0.03$)		\\
$\sigma$	& 0.05	&	0.04					&	0.04					&	0.05					&	0.05					&	0.05					&	0.05					\\

\hline
	\end{tabu}
  \label{photL107}
  \end{center}
\end{table}

\begin{table}[H]
  \begin{center}
  \caption{Photometric coefficients for NGC5904}
  \setlength{\extrarowheight}{5pt}
	\setlength{\tabcolsep}{5pt}
	\begin{tabu}{c|c|cc|cc|cc}
	\hline
	& \multicolumn{1}{c|}{B} & \multicolumn{2}{c|}{V} & \multicolumn{2}{c|}{R} & \multicolumn{2}{c}{I} \\
	\hline
	
N	&	787			& \multicolumn{2}{c|}{1162} & \multicolumn{2}{c|}{1574} & \multicolumn{2}{c}{1278} \\
S	&	0.993					&	1.010					&	1.010					&	1.004					&	1.007					&	0.997					&	1.000					\\
	&	($\pm0.003$)	&	($\pm0.002$)	&	($\pm0.002$)	&	($\pm0.002$)	&	($\pm0.002$)	&	($\pm0.002$)	&	($\pm0.002$)	\\
Z	&	-6.42					&	-6.90					&	-6.90					&	-6.72					&	-6.76					&	-5.99					&	-6.03					\\
	&	($\pm0.05$)		&	($\pm0.04$)		&	($\pm0.04$)		&	($\pm0.04$)		&	($\pm0.04$)		&	($\pm0.04$)		&	($\pm0.03$)		\\
$\sigma$	&	0.07	&	0.06					&	0.06					&	0.08					&	0.08					&	0.07					&	0.06					\\

\hline
	\end{tabu}
  \label{photNGC5904}
  \end{center}
\end{table}

\begin{table}[H]
  \begin{center}
  \caption{Photometric coefficients for NGC6341}
  \setlength{\extrarowheight}{5pt}
	\setlength{\tabcolsep}{5pt}
	\begin{tabu}{c|c|cc|cc|cc}
	\hline
	& \multicolumn{1}{c|}{B} & \multicolumn{2}{c|}{V} & \multicolumn{2}{c|}{R} & \multicolumn{2}{c}{I} \\
	\hline

N	& 670	& \multicolumn{2}{c|}{1615}	& \multicolumn{2}{c|}{1392}	& \multicolumn{2}{c}{1496} 		\\
S	&	0.996					& 1.007 				&	1.007					&	1.003						&	1.006					&	0.994					& 0.995					\\
  &	($\pm0.007$)	& ($\pm0.002$)	&	($\pm0.002$)	&	($\pm0.002$)		&	($\pm0.002$)	&	($\pm0.002$)	& ($\pm0.002$)	\\
Z	&	-6.54					& -6.92   			& -6.92 				& -6.72  					& -6.77					& -5.99  				& -6.00  				\\
  & ($\pm0.12$)		& ($\pm0.03$) 	& ($\pm0.03$) 	& ($\pm0.03$) 		& ($\pm0.03$) 	& ($\pm0.03$) 	& ($\pm0.03$) 	\\
$\sigma$ & 0.16   & 0.07 					& 0.07					& 0.08						& 0.08					& 0.07 					& 0.06 					\\

\hline
	\end{tabu}
  \label{photNGC6341}
  \end{center}
\end{table}

\section{Photometry of small bodies observations}

Photometry of small bodies in the Solar System always faces the problem of finding reliable standards in the observed field. Usually in-field photometric standards are missing, and one way to obtain the magnitudes is to apply relative photometry. In this method it is essential to have significant number of stars in the observed field, which will be used as secondary standards. The problem with the so called 'secondary standards' is that they actually are not standard stars. Many of them suffer from numerous disadvantages: they are variable, they are not stars, magnitudes are not accurate, etc. Hence their number in the process of calibration is so important. Using a large number of secondary standards can minimise the impact of biased instrumental magnitudes. Fortunately observations with a wide-field telescope system (field of view $>1^{\circ}$), offer thousands of stars with suitable S/N ratio in the field of every target. The significant number of stars which can be used as secondary standards makes the calibration of the instrumental magnitudes possible. As stated in the Introduction the most suitable source for a large number of stars homogeneously distributed all over the sky is PS1. The equations (\ref{eq:bggr})$\div$(\ref{eq:iiri}), can be used to transform \textit{gri} to BVRI, making the data compatible with any other observation obtained with or transformed to this standard system.

The procedure, that we followed to calibrate observations of small bodies, follow the steps of the verification of the transformation equations, described in the previous section. These steps are:

\begin{itemize}
\renewcommand{\labelitemi}{$\bullet$}
			\item{Find initial astrometric solution of the observed field.}
			\item{Identify all point sources in the image and record their in-image position coordinates (X, Y), along with their celestial coordinates ($\alpha, \delta$).}
			\item{Use the PS1 catalogue search tool\textsuperscript{\ref{PS1DB}} to cross-identify all selected stars.}
			\item{Derive the instrumental magnitudes of all cross-identified stars.}
			\item{Use the transformation equations (\ref{eq:bggr})$\div$(\ref{eq:iiri}) with their coefficients (Table \ref{equation_results}) to transform all \textit{gri} to $B^\prime V^\prime R^\prime I^\prime$ magnitudes.}
			\item{Derive the photometric coefficients, $Z$ and $S$, of each image, comparing the instrumental magnitudes with $B^\prime V^\prime R^\prime I^\prime$ magnitudes.}
			\item{With the derived photometric zero point and slope calibrate the instrumental magnitudes of the observed object (comet or asteroid) to standard BVRI magnitude.}
\end{itemize}

To optimise the procedure, and to provide robust transformations, it is advisable  to set some criteria for sampling reasonable sets of stars, which will be used in deriving the transformation coefficients of the images. One possibility in this respect is to use the astrometric and photometric criteria, described in this paper, in order to minimise the instrumental magnitudes uncertainty caused by poor photometry/astrometry.

To give an example how the described algorithm works, we checked the BVRI images of NGC5904 for moving objects using the software package \textbf{ASTROMETRICA}\footnote{\url{http://www.astrometrica.at/}} by Raab (2012). One of the detected objects was the main-belt asteroid $48220$. Application of the procedure to that object, with unknown photometric parameters, resulted in following calculated magnitudes: $B\!=\!19.12\pm0.14$, $V\!=\!18.29\pm0.08$, $R\!=\!17.93\pm0.08$ and $I\!=\!17.61\pm0.11$. The V magnitude is in good agreement with the ephemeris magnitude given by \textbf{HORIZONS\footnote{\url{https://ssd.jpl.nasa.gov/horizons.cgi}\label{HEG}}}, the on-line Solar system data and ephemeris described by Giorgini et al. (1996). The derived colours of the asteroid are $B-V=0.829\pm0.165$, $V-R=0.364\pm0.109$ and $R-I=0.317\pm0.131$. Within the limits of the photometric uncertainties, these colours resemble closely the colours of the Sun (Ram\'{\i}rez et al. 2012). This neutral reflectance is a typical property of carbonaceous bodies, which gives a reason to consider $48220$ a member of the mostly populated taxonomic class of C-type asteroids.

\section{Conclusions}

In this paper we present relations between two standard star systems, namely Stetson’s BVRI and Pan-STARRS1 \textit{gri}, established by means of seven transformation equations. Using cross-identification, a set of 15074 stars form Stetson's database\textsuperscript{\ref{SDB}} and PS1 catalogue\textsuperscript{\ref{PS1DB}} was extracted and used to derive the transformation equations coefficients. Comparison of the derived coefficients and their uncertainties revealed that the linear fit gives reliable results. Verification tests were performed applying them to three Stetson's standard fields. The results of these tests confirmed the reliability of the transformation equations. There were several features (like the slight deviation from linearity for extremely red stars, or the influence of Population II stars on the results for NGC6341) that do not impact the final results, but can be subject of a future study, going into more details in that direction. The proposed procedure of relative photometry can be used for wide range of tasks (when standards do not exist in the observed field): observations of variable stars, exoplanets, optical transients etc. The relative photometry of small bodies observations using these transformations is a possible task, and this is the field where they will find wide and effective application. As a first application of the transformation equations, on a small body in the Solar system, we derived the magnitudes and colours of the minor planet 48220.

\section*{Acknowledgements}

The Pan-STARRS1 Surveys (PS1) and the PS1 public science archive have been made possible through contributions by the Institute for Astronomy, the University of Hawaii, the Pan-STARRS Project Office, the Max-Planck Society and its participating institutes, the Max Planck Institute for Astronomy, Heidelberg and the Max Planck Institute for Extraterrestrial Physics, Garching, The Johns Hopkins University, Durham University, the University of Edinburgh, the Queen's University Belfast, the Harvard-Smithsonian Center for Astrophysics, the Las Cumbres Observatory Global Telescope Network Incorporated, the National Central University of Taiwan, the Space Telescope Science Institute, the National Aeronautics and Space Administration under Grant No. NNX08AR22G issued through the Planetary Science Division of the NASA Science Mission Directorate, the National Science Foundation Grant No. AST-1238877, the University of Maryland, Eotvos Lorand University (ELTE), the Los Alamos National Laboratory, and the Gordon and Betty Moore Foundation.

This research was partially supported by the Bulgarian National Science Fund of the Ministry of Education and Science under the following grants: DN 08-1/2016 and DO 02-362.

\end{document}